\documentclass{jfm}
\usepackage{graphicx}
\usepackage{epstopdf, epsfig}
\usepackage{xcolor} 
\usepackage{amsmath,amsbsy}

\shorttitle{The role of monolayer viscosity in Langmuir film hole closure dynamics}
\shortauthor{L. L. Jia and M. J. Shelley}

\title{The role of monolayer viscosity in Langmuir film hole closure dynamics}

\author{Leroy L. Jia\aff{1}
  \corresp{\email{ljia@flatironinstitute.org}}
 \and Michael J. Shelley\aff{1}\aff{2}}

\affiliation{\aff{1}Center for Computational Biology, Flatiron Institute, New York, NY 10010, USA
\aff{2} Courant Institute of Mathematical Sciences, New York University, New York, NY 10012, USA}

\begin{document}

\maketitle
\begin{abstract}
We re-examine the model proposed by \cite{Alexander_etal2006} for the closing of a circular hole in a molecularly thin incompressible Langmuir film
situated on a Stokesian subfluid. For simplicity their model assumes that the surface phase is inviscid which leads to the result that the cavity area decreases at a constant rate determined by the ratio of edge tension to subfluid viscosity. We reformulate the problem, allowing for a regularizing monolayer viscosity. The viscosity-dependent corrections to the hole dynamics are analyzed and found to be nontrivial, even when the monolayer viscosity is small; these corrections may explain the departure of experimental data from the theoretical prediction when the hole radius becomes comparable to the Saffman-Delbr\"uck length. Through fitting, we find the edge tension could be as much as eight times larger ($\sim 5.5$ pN) than previously reported under these relaxed assumptions. 
\end{abstract}

\section{Introduction}
An understanding of the hydrodynamics of a quasi-2D fluid interface coupled to a bulk 3D subphase is crucial to modeling the dynamics of a wide variety of biological systems such as lipid membranes~(\cite{StoneMcConnell1995}), bacterial biofilms~(\cite{Petroff_etal2005}), and algae colonies~(\cite{Drescher_etal2009}). One notable theoretical and experimental study of such a system was conducted by~\cite{Alexander_etal2006}, who examined the tension-driven closing of a cavity punctured in a PDMS monolayer situated on a flat water-air interface. 
The authors derived an analytical solution for the axisymmetric fluid motion, predicting that the area of the cavity closes at a fixed rate that depends only on the ratio of monolayer line tension to bulk viscosity; by fitting experimental data in a linear regime of the area vs. time plot, they obtained an estimate for the line tension. The same authors later reapplied this model along with improved experimental methods to obtain a more accurate value for the monolayer line tension, which they reported to be $0.69\pm 0.02$ pN in~\cite{Zou_etal2010}.

The reduction of this problem to the fitting of a single parameter hinges on a crucial assumption: the monolayer viscosity is vanishingly small. While there is experimental evidence supporting the claim that the film dynamics is dominated by subphase viscosity rather than monolayer viscosity~(\cite{Mann_etal1995}), and while there are strongly linear regimes in the experimental area vs. time data in both~\cite{Alexander_etal2006} and~\cite{Zou_etal2010}, the data as a whole is visibly nonlinear, especially at later times when the hole size is small. This indicates that the inviscid assumption may not hold as well as previously thought and that viscous effects of the monolayer may be playing a role in the closure process.

Here we revisit the theoretical model proposed by~\cite{Alexander_etal2006} and amend it by considering the effects of a regularizing monolayer viscosity. This simple change preserves much of the structure of the problem but introduces some notable differences. We find that while the bulk equations are unchanged in the axisymmetric case, the boundary conditions can depend strongly on this monolayer viscosity, especially in regimes where the cavity radius is small relative to the Saffman-Delbr\"uck length--that is, the ratio of the 2D monolayer and 3D bulk viscosities.
In this regime, the predicted dependence of cavity area on time becomes noticeably nonlinear, 
which impacts the fitting of experimental data and consequently suggests that these previous studies have underestimated the edge tension by roughly one order of magnitude. 

\section{Summary of the original model}

We begin by reviewing the system description given in~\cite{Alexander_etal2006}. Consider a molecularly thin (so essentially two-dimensional) fluid monolayer situated on a three-dimensional half-space of Stokes fluid with flow field $\boldsymbol{u}$ and viscosity $\mu$.  The domain of the monolayer is taken to be the annulus centered at the origin in the $z=0$ plane, $\mathcal{D} = \{(r,0): 0< R_i< r < R_o\}$, where $R_o$ can be infinite. The interface is assumed to be planar for all times, and the flow field of the interface, $\boldsymbol{U} = \boldsymbol{u}(z=0)$, is assumed to have no $z$-component. If the monolayer is 2D incompressible, then the radial component of $\boldsymbol{U}$, which we denote by  $U(r)$, satisfies
\begin{equation}
    \frac{1}{r}\frac{\mathrm{d}}{\mathrm{d}r} (rU)= 0
    \label{divUeqn}
\end{equation}
so that it necessarily has the form
\begin{equation}
U(r) = \frac{F}{r}
\label{Ueqn}
\end{equation}
for some constant $F$ when $R_i <r <R_o$. In the absence of tangential stresses, the tangential component of $\boldsymbol{U}$ vanishes.

In deriving the momentum equation,~\cite{Alexander_etal2006} employ the so-called ``hydrostatic approximation'': the monolayer viscosity is so small that the film may be treated as inviscid. Hence, it is simply the pressure gradient that balances the shear forces from the fluid subphase where it is in contact with the monolayer. Elsewhere, a stress-free interface is assumed:
\begin{equation}
\left.\mu \frac{\partial u}{\partial z}\right|_{z=0} = 0\text{ , } 0<r<R_i
\label{momeqn1}
\end{equation}
\begin{equation}
\left.\mu \frac{\partial u}{\partial z}\right|_{z=0} = -\frac{\mathrm{d}P}{\mathrm{d}r}\text{ , } R_i<r<R_o
\label{momeqn2}
\end{equation}
\begin{equation}
\left.\mu \frac{\partial u}{\partial z}\right|_{z=0} = 0\text{ , } r>R_o
\label{momeqn3}
\end{equation}
where $u$ is the radial component of $\boldsymbol{u}$ and $P$ is the 2D pressure inside of the monolayer. Using standard Hankel transform methods, an equivalent set of triple integral equations can be derived:
\begin{equation}
\int_0^\infty \mathrm{d}k~ k b(k) J_1(kr) = 0\text{ , } 0<r<R_i
\label{tie1}
\end{equation}
\begin{equation}
\int_0^\infty \mathrm{d}k~ b(k) J_1(kr) =  \frac{\mu F}{r} \text{, }R_i<r<R_o
\label{tie2}
\end{equation}
\begin{equation}
\int_0^\infty \mathrm{d}k~ k b(k) J_1(kr) = 0\text{ , } r>R_o
\label{tie3}
\end{equation}
where we have defined $k$ to be the wavenumber and 
\begin{equation}
b(k) = \int_0^\infty \mathrm{d}r~ r  \mu \frac{\partial u}{\partial z}(r,z=0) J_1(kr).
\label{bdef}
\end{equation}



The problem is then closed by enforcing the boundary conditions of normal stress balance. In this case, these are simply  the Young-Laplace relations
\begin{equation}
-P_i^- = -P_i^+  - \frac{\gamma}{R_i}
\label{YLeqn1}
\end{equation}
\begin{equation}
-P_o^+ = -P_o^- + \frac{\gamma}{R_o}
\label{YLeqn2}
\end{equation}
where the superscript plus and minus represent limits from the right and left, respectively. Since the cavity does not contain a significant concentration of molecules, the pressures inside of the cavity and at infinity are negligible, and Eqns.~(\ref{YLeqn1}) and~(\ref{YLeqn2}) can be combined into a single equation for the internal pressure difference at the interfaces,
\begin{equation}
P_o^- -P_i^+ = \gamma \left(\frac{1}{R_o} + \frac{1}{R_i}\right).
\label{YLeqn3}
\end{equation}
The kinematic boundary condition relates the change in radius with respect to time $t$ and the fluid velocity at the boundary:
\begin{equation}
\frac{\mathrm{d}R_i}{\mathrm{d}t} =U(R_i) = \frac{F}{R_i}.
\label{kbc}
\end{equation}
The problem is readily solved by converting the triple integral equations into a singular integral equation and numerically inverting to solve for the shear stress at each time step; for details the reader is referred to~\cite{Jia_etal2021}. However, in the limit of $R_o\to \infty$, the triple integral equations Eqns.~(\ref{tie1})-(\ref{tie3}) admit a simple analytical solution
\begin{equation}
b(k) = \mu F\frac{\sin k R_i}{k R_i}.
\label{bsoln}
\end{equation}
Employing the Hankel inversion theorem and substituting into the momentum equation yields  (cf. Formula~(6.693.1) of~\cite{GradshteynRyzhik2007}, differentiate with respect to $\alpha$, let $\nu \to 0$, and take a negative sign)
\begin{equation}
-\frac{\mathrm{d}P}{\mathrm{d}r} = \mu F\int_0^\infty \mathrm{d}k~ k \frac{\sin kR_i}{kR_i }J_1(kr) = \frac{\mu F}{r\sqrt{r^2-R_i^2}} \chi(r>R_i), 
\end{equation}
where $\chi$ is a characteristic function. The monolayer pressure is then
\begin{equation}
P =- \mu F\frac{\cos^{-1} (R_i/r)}{R_i}
\label{Peqn}
\end{equation}
where the constant of integration has been taken to be zero. This gives
\begin{equation}
P_o^- - P_i^+ = -\frac{\pi \mu F}{2R_i},
\end{equation}
which, when combined with Eqn.~(\ref{YLeqn3}), produces the simple result 
\begin{equation}
F =- \frac{2 \gamma}{\pi\mu}.
\label{Fresult}
\end{equation}
Since $dA/dt = 2\pi F$, the inviscid model predicts that the hole simply closes via ``flow by curvature'': the area of the cavity decreases linearly at a rate independent of  $R_i$ until it fully closes. As a consequence, the radial surface velocity at the interface $U(R_i) =F/R_i$ diverges as the hole radius goes to zero in this model.

For completeness, we remark that if $0<r<R_i$, 
\begin{equation}
    U(r) = \frac{1}{\mu}\int_0^\infty \mathrm{d}k~b(k)J_1(kr) = \frac{F}{r}\left[1-\sqrt{1 - \left(\frac{r}{R_i}\right)^2}\right]
\end{equation}
gives the surface velocity inside of the cavity. As expected, the flow is not incompressible in this region.

\section{Adjusting the model}

The formulation as presented rests on the assumption that the film is surface inviscid 
and therefore the surface shear stress is in  equilibrium with the surface pressure inside $\mathcal{D}$
for all times. We will amend the above treatment by allowing for the possibility of a monolayer viscosity. We now model the surface phase as governed by the incompressible two-dimensional Stokes equations, again forced by the surface shear stress of the bulk phase. In polar coordinates for an axisymmetric system these equations are:
\begin{equation}
\frac{1}{r} \frac{\mathrm{d}}{\mathrm{d}r}(rU) = 0
\label{newdivfree}
\end{equation}
\begin{equation}
\left. \mu \frac{\partial u}{\partial z}\right|_{z=0} = \chi(R_i<r<R_o) \left(-\frac{\mathrm{d}P}{\mathrm{d}r} + \eta \mathcal{L}[U]\right),
\label{momeqn}
\end{equation}
where $\eta$ is the monolayer viscosity and $\mathcal{L} = \partial_{rr} + (1/r)\partial_r -1/r^2$. As before, Eqn.~(\ref{newdivfree}) implies $U=F/r$ for some constant $F$. Amending Eqns.~(\ref{YLeqn1}) and (\ref{YLeqn2}), the stress balance at the interfaces becomes
\begin{equation}
-P_i^- = -P_i^+ -2\eta \frac{F}{R_i^2} - \frac{\gamma}{R_i}
\label{YLeqn1new}
\end{equation}
\begin{equation}
-P_o^+ = -P_o^- -2\eta \frac{F}{R_o^2} + \frac{\gamma}{R_o}.
\label{YLeqn2new}
\end{equation}
The form of the kinematic boundary condition Eqn.~(\ref{kbc}) is unchanged. In keeping with~\cite{Alexander_etal2006}, we will focus on the $R_o\to\infty$ limit with $P_i^-=P_o^+ = 0$.

It is instructive to consider the dimensionless version of the above equations. We choose $R_i$ to be the length scale, $V=\gamma/(\mu R_i)$ to be the velocity scale, and $\mu V$ to be the pressure scale. Identifying dimensionless quantities with their dimensional counterparts, the resulting momentum balance is
\begin{equation}
    \left. \frac{\partial u}{\partial z}\right|_{z=0} = \chi({r}>R_i)\left(-\frac{\mathrm{d}P}{\mathrm{d}r} + \frac{\eta/\mu}{R_i} \mathcal{L}[U]\right)
\end{equation}
which gives the dimensionless parameter $\beta = \ell_{\mathrm{SD}}/R_i$; here, $\ell_{\mathrm{SD}} = \eta/\mu$ is the Saffman-Delbr\"uck length for membranes. The corresponding dimensionless stress balance at the inner radius is
\begin{equation}
    -P_i^+ - 2 \beta F =1.
\end{equation}
We now discriminate between two subtle yet distinct cases. If $\beta = 0$, the monolayer viscosity drops out of the problem completely, and the solution follows the prescription of~\cite{Alexander_etal2006} outlined  above. The constant $F$ necessarily scales like $VR_i$, presaging the general form of Eqn.~(\ref{Fresult}).
On the other hand, when $\beta >0$, the monolayer viscosity continues to play no role in the bulk. This is because the radial annular flow $U=F/r$ is both incompressible and irrotational so that $\mathcal{L}[U]=0$.
Thus, regardless of $\beta$,  Eqns.~(\ref{momeqn1})-(\ref{momeqn3}) hold, so that Eqns.~(\ref{bsoln}) and (\ref{Peqn}) for $b(k)$ and $P$ are common to both scenarios. The effects of $\beta >0$ only materialize at the boundary, where an additional viscosity term emerges, and become non-negligible as $R_i$ goes to zero.

\begin{figure}
\centering
\includegraphics[scale=0.165]{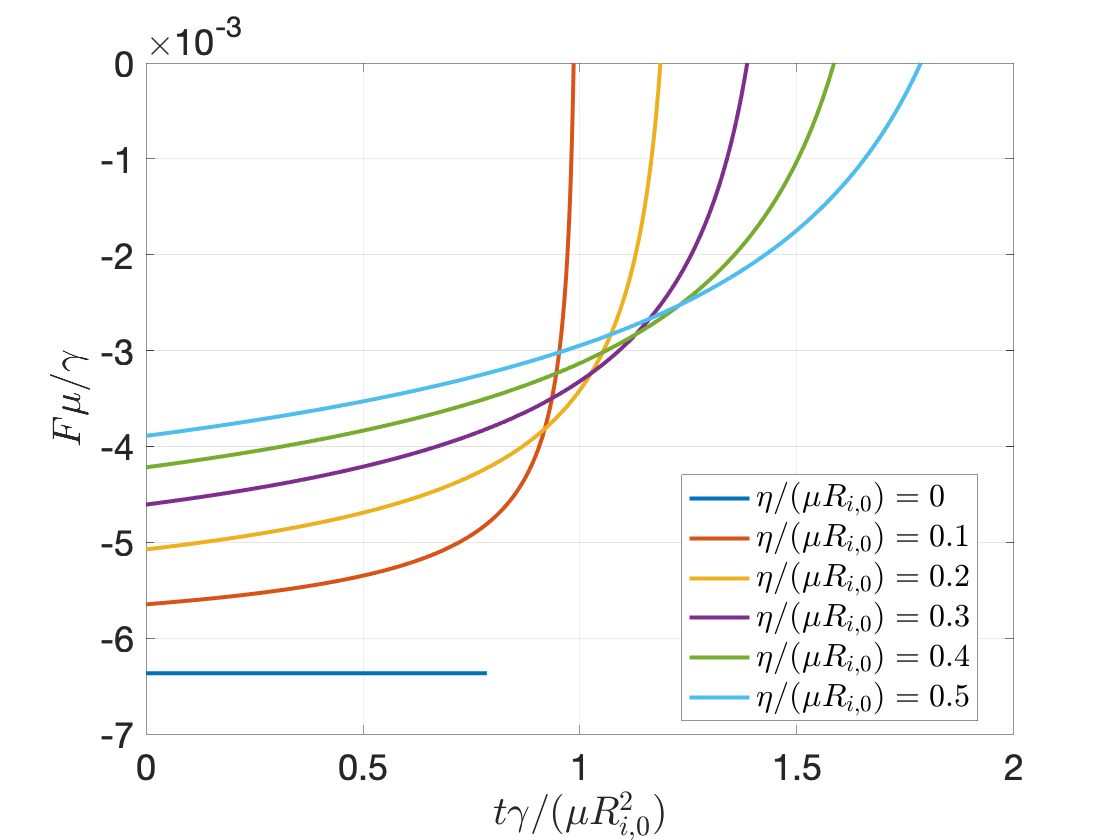}
\includegraphics[scale=0.165]{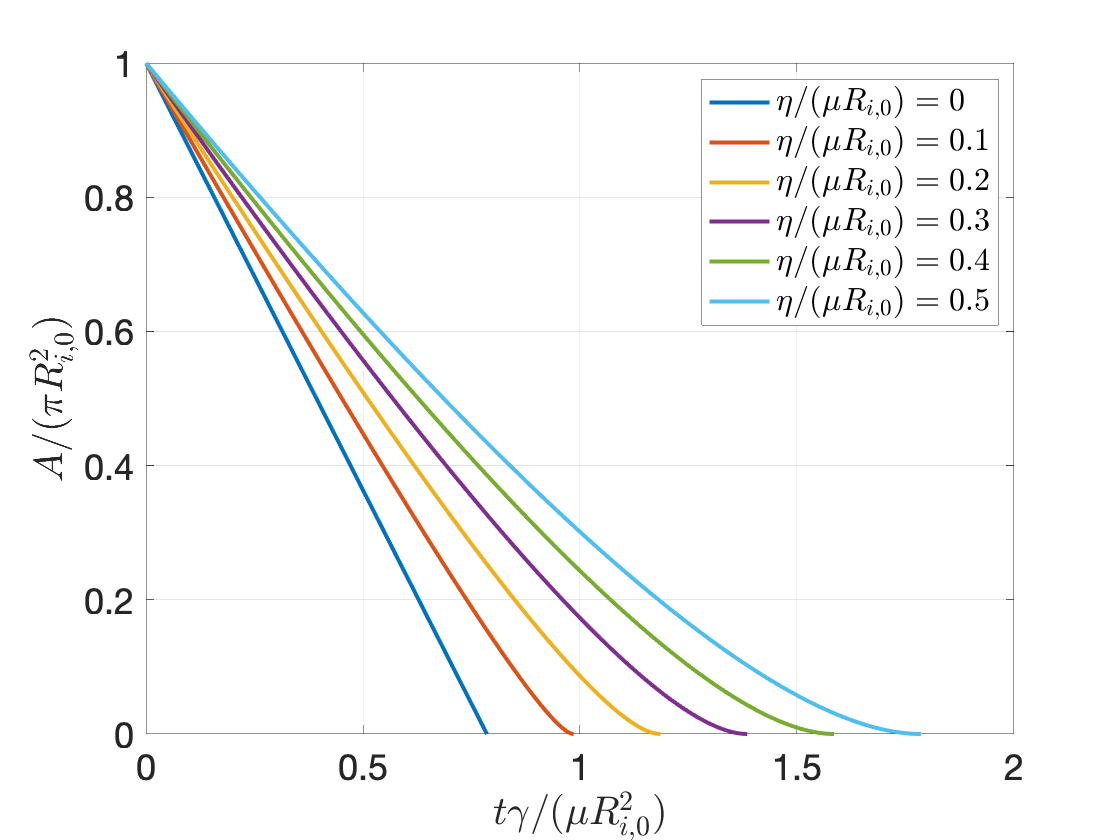}
\caption{
 (a) The constant $F$ from Eqn.~(\ref{newFeqn}), which determines the rate of decrease of the cavity area, as a function of time for different values of the parameter $\beta=\eta/(\mu R_{i,0})$, with $R_{i,0}$ the initial cavity radius. Parameters: $\gamma = 0.01$, $\mu = 10$. Note that $F(t^*)$ is not zero when $\eta=0$, while $F$ increases to zero linearly with a slope proportional to $\eta^{-2}$ when $\eta>0$ and the hole radius is small compared to the Saffman-Delbr\"uck length $\ell_{\mathrm{SD}}=\eta/\mu$. (b) Increasing monolayer viscosity slows down the closing of the cavity. Same parameters. The cavity area exhibits an initial linear small $\beta$ regime followed by a quadratic large $\beta$ regime as the cavity radius becomes comparable to $\ell_{\mathrm{SD}}$.} 
\label{Fplots}
\end{figure}

Taking the difference of Eqns.~(\ref{YLeqn1new}) and (\ref{YLeqn2new}), setting $P_i^- = P_o^+=0$ as before, and using Eqn.~(\ref{Peqn}) for $P$ gives the corrected
\begin{equation}
F = -\frac{\gamma}{ (2\eta/R_i )+ (\pi\mu/2)}.
\label{newFeqn}
\end{equation}
Note that $\eta>0$ regularizes $F$ so that $F\to 0$ as $R_i\to 0$, whereas $F$ is independent of $R_i$ in the $\beta=0$ surface inviscid case (Fig.~\ref{Fplots}a). The kinematic boundary condition can then be integrated to obtain an explicit relation for the cavity area $A$ as a function of time
\begin{equation}
A(t) = \frac{4}{\pi \mu^2} \left[2\eta - \sqrt{4\eta^2 + \pi \gamma \mu(t-t^*) }\right]^2,
\label{Avstnew}
\end{equation}
where $t^* = (2\eta R_{i,0} + \pi\mu R_{i,0}^2/4 )/\gamma$ is the closing time of the cavity written in terms of the initial radius $R_{i,0}$. If this result is substituted back into Eqn.~(\ref{newFeqn}), we find an explicit relation for $F$ as a function of time:
\begin{equation}
    F(t) = - \frac{2\gamma}{\pi \mu} \left(1 - \frac{1}{\sqrt{1 + \pi \mu \gamma (t^*-t)/(4\eta^2)}}\right)
\end{equation}
If $\eta \to 0$ in this expression, we recover Eqn.~(\ref{Fresult}). On the other hand, fixing $\eta$ and expanding in small $t^*-t$ gives the result
\begin{equation}
    F(t) = -\frac{2\gamma}{\pi \mu}\left[ \frac{\pi \mu \gamma}{8\eta^2} (t^*-t) + \hdots\right] = -\frac{\gamma^2}{4\eta^2} (t^*-t) + \hdots
\end{equation}
which demonstrates the rapid variation of $F$ to zero as the cavity closes in the  case of a viscous monolayer. 

The surface inviscid result of linearly decreasing area found by~\cite{Alexander_etal2006} is a valid approximation for early times if monolayer viscosity is indeed small, as Fig.~\ref{Fplots}a illustrates. However, the predicted slope of $-4\gamma/\mu$ does not hold as well for intermediate times in the presence of nonzero viscosity.
The $O(\beta)$ perturbation to the slope at early times can be found by calculating
\begin{equation}
    \frac{\mathrm{d}A}{\mathrm{d}t}|_{t=0} = 2\pi F|_{t=0} = -\frac{4\gamma}{\mu} \left(1 - \frac{2\eta}{\sqrt{\pi \mu \gamma t^*}} + \hdots\right)
\end{equation}
where we have taken $\eta$ sufficiently smaller than $\mu R_i$. Thus, the line tension value obtained from fitting data to the inviscid model is likely an underestimate.

\begin{figure}
    \centering
    \includegraphics[scale=0.179]{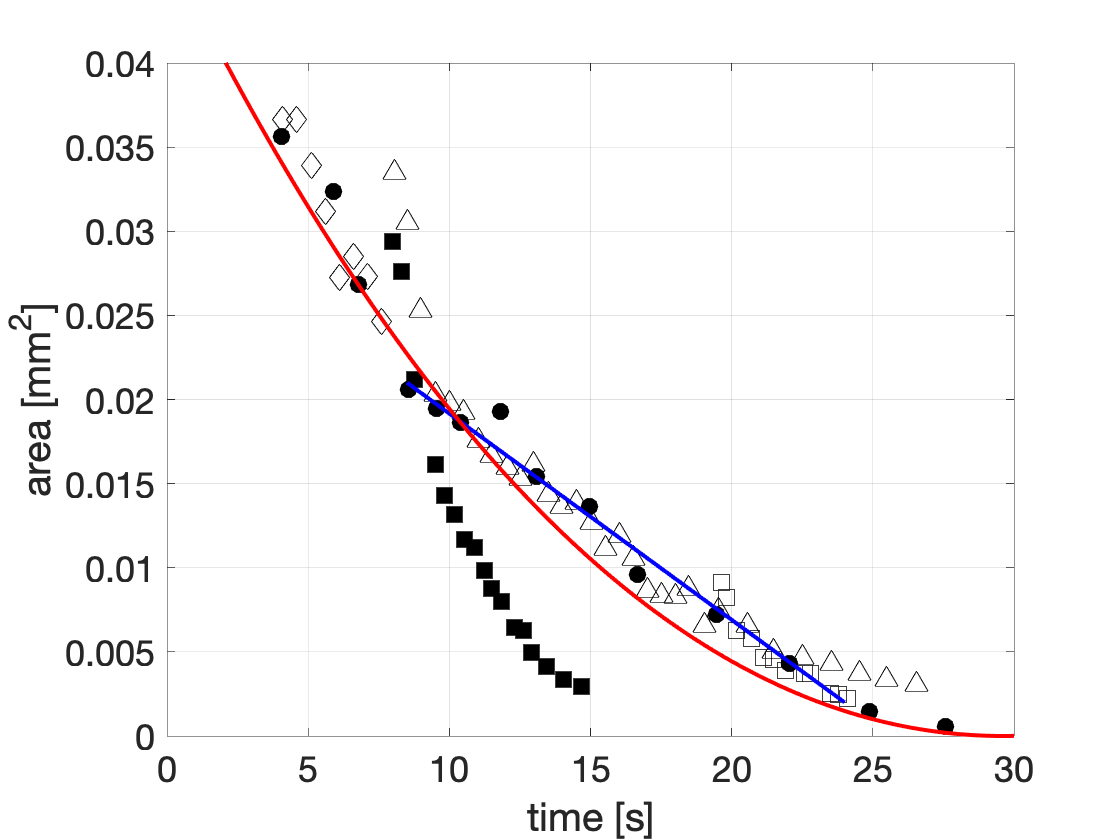}
    \includegraphics[scale=0.204]{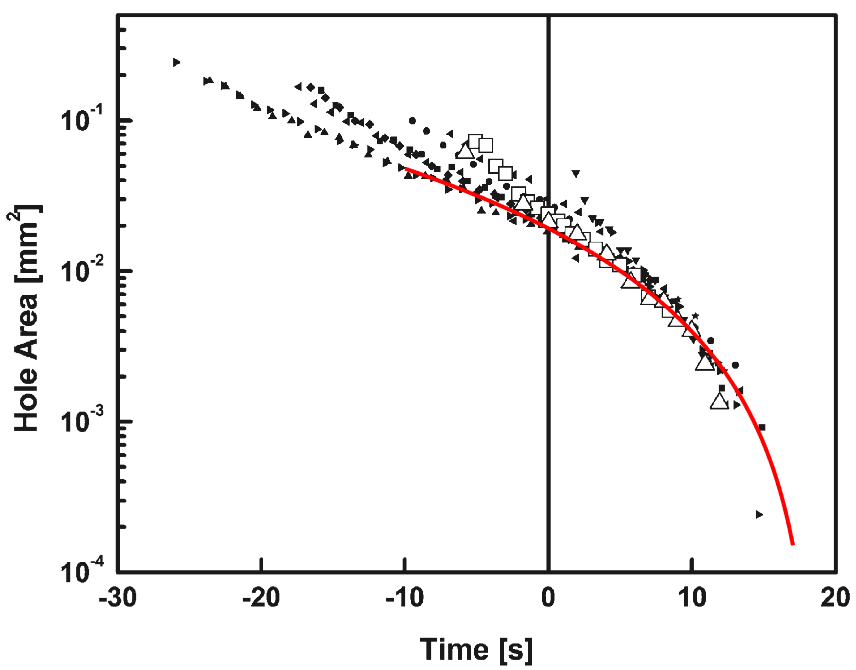}
    \caption{(a) A reproduction of Figure 6 of~\cite{Alexander_etal2006} showing five different runs of experimental hole sizes as a function of time. As in the original paper, each run has been shifted horizontally to maximize overlap. The data are nonlinear when the hole becomes small, which could possibly be an effect of monolayer viscosity. The blue line is the linear fit that appeared in the paper originally; the red line is the fit of Eqn.~(23) which uses monolayer viscosity as an additional fit parameter and more accurately describes a broader range of data. The line tension and monolayer viscosity obtained from fitting are $\gamma = 5.54\times 10^{-12}$ N and $\eta = 6.80\times 10^{-7}$ kg/s.  {(b) A reproduction of Figure 4 from~\cite{Zou_etal2010}, plotting the fit with the same parameters with a time shift on top of the experimental data. The vertical axis is on a base 10 log scale.}}
    \label{newfitplot}
\end{figure}

In the opposite regime where the hole becomes sufficiently small, the viscous stress dominates, and $F \sim -\gamma R_i/(2\eta) $. That is, the cavity area changes  quadratically with time in this regime, and the inviscid monolayer model consequently underestimates the closure time.
The experimental data in Figure 6 of~\cite{Alexander_etal2006} 
demonstrate a prominent quadratic-like slowing when the hole is small that may be representative of a monolayer viscosity. Figure~\ref{newfitplot}a compares the authors' original linear fit to the result of fitting to the modified area vs. time relation Eqn.~(\ref{Avstnew}); the modification allows for a more accurate description of a broader range of data. The fit parameters were found to be a line tension of $\gamma = 5.54\times 10^{-12}$ N and monolayer viscosity of $\eta = 6.80\times 10^{-7}$ kg/s, for the value of $\mu = 1.063\times 10^{-3}$ kg/m$\text{ s}$ used in~\cite{Alexander_etal2006}. The corresponding Saffman-Delbr\"uck length is $\ell_{\mathrm{SD}} = 0.64$ mm, which gives a $\beta$ that is $O(1)$ at the initial time and confirms that the layer should not be treated as inviscid.

Figure~\ref{newfitplot}b compares the same fit with the more recent experimental data from~\cite{Zou_etal2010} with a horizontal shift selected so that there is maximum overlap (the data in the original image were also shifted in this manner); again, we see a strong agreement between the modified theory and the experiment.

\section{Conclusion}

Effectively, two regimes comprise the closing of a (sufficiently large) cavity in a Langmuir film. The first is a tension-dominated regime where the area decreases linearly. This is followed by a viscosity-dominated regime where the area decreases quadratically just as the hole is about to close, effectively regularizing the singularity. The two regimes can be delineated by the dimensionless parameter $\beta$, which is a ratio of the Saffman-Delbr\"uck length to the cavity radius.

We applied our theory to the experiments of~\cite{Alexander_etal2006}. By allowing for a monolayer viscosity in the model they proposed for the closing of a Langmuir film, the improved theory matches the original experiment more closely over a broader range of data. From fitting, we find a line tension of approximately 5.5 pN, which is the same order of magnitude as the previously reported values of $0.69\pm 0.02$ pN from~\cite{Zou_etal2010} and $1.1\pm 0.3$ pN from~\cite{Mann_etal1992}, but is different enough to see that the effects of monolayer viscosity are perhaps not as negligible for this system as the authors believed.


As mentioned above, the vanishing of $\eta$ from the bulk equations is a consequence of the symmetries of this particular problem. The effects of monolayer viscosity in a nonannular domain should be investigated as well. The flow field of a disk-shaped monolayer of actively rotating colloidal magnets has already produced one example of an axisymmetric system where monolayer viscosity plays a nontrivial role in the bulk~(\cite{Jia_etal2021}); presumably there are many others. It would be interesting to study the effects of a surface viscosity on domain relaxation in nonaxisymmetric Langmuir films, akin to the work done by~\cite{Alexander_etal2007}.

{\bf Acknowledgements:}
MJS acknowledges support by the National Science Foundation under awards DMR-1420073 (NYU MRSEC) and DMR-2004469. 




\bibliographystyle{jfm}
\bibliography{lfrefs}








\end{document}